\documentclass[12pt,fleqn]{article}

\usepackage{graphicx}
\usepackage{pdftricks}
\begin{document}

\title{\Large Thermodynamic $R$-diagrams reveal solid-like fluid states}

\author{George Ruppeiner\footnote{New College of Florida, Sarasota, Florida, 34243, USA (ruppeiner@ncf.edu)}, Peter Mausbach\footnote{Cologne University of Applied Sciences, Cologne, D-50679, Germany} and Helge-Otmar May\footnote{University of Applied Sciences, Darmstadt, D-64295, Germany}}

\maketitle

\begin{abstract}
We evaluate the thermodynamic curvature $R$ for fluid argon, hydrogen, carbon dioxide, and water. For these fluids, $R$ is mostly negative, but we also find significant regimes of positive $R$, which we interpret as indicating solid-like fluid properties. Regimes of positive $R$ are present in all four fluids at very high pressure. Water has, in addition, a narrow slab of positive $R$ in the stable liquid phase near its triple point. Also, water is the only fluid we found having $R$ decrease on cooling into the metastable liquid phase, consistent with a possible second critical point.
\end{abstract}

\\

\par
Extracting general information from thermodynamic properties poses a challenge. Which properties to feature depends strongly on the type of system under consideration, the regime of interest, and the broader context of applicable models from statistical mechanics. Very useful for comparison between different systems would be some representation not dependent on such subjective features. In this paper, we propose $R$-diagrams, based on the thermodynamic curvature $R$, as such a representation. At a glance, $R$-diagrams reveal a number of important fluid properties. We key here particularly on locating solid-like properties of fluids, with particular attention to anomalies in liquid water in the vicinity of its triple point.

\section{Introduction}

Thermodynamic curvature is an element of thermodynamic metric geometry. A pioneering paper on this was by Weinhold \cite{Weinhold1975} who introduced a thermodynamic energy inner product. This led to the work of Ruppeiner \cite{Ruppeiner1979} who wrote a Riemannian thermodynamic entropy metric to represent thermodynamic fluctuation theory, and was the first to systematically calculate the thermodynamic Ricci curvature scalar $R$. A parallel effort was by Andresen, Salamon, and Berry \cite{Andresen1984} who began the systematic application of the thermodynamic entropy metric to finite-time thermodynamic processes.

\par
Thermodynamic fluctuation theory \cite{Landau} is essential to the discussion of water's anomalies \cite{Debenedetti2003b}, for example, and $R$ is an inevitable element of this theory \cite{Ruppeiner1995}. $R$ is a direct measure of the size of organized mesoscopic fluctuating structures in thermodynamic systems. For a recent review of $R$ in a broader context, including black hole thermodynamics, see \cite{Ruppeiner2014}. Other recent evaluations of $R$ are \cite{Mirza2013,Ubriaco2013}.

\par
In this paper, we key on the sign of $R$, which appears to be connected with the character of interparticle interactions: $R>0$ if repulsive interactions dominate, as in solids, and $R<0$ if attractive interactions dominate, as near the critical point \cite{Ruppeiner2010}. The first indication of a repulsive/attractive sign interpretation for $R$ came from the quantum Fermi/Bose gasses, with uniformly positive/negative $R$ values (in the $R$ sign convention used here\footnote{Our sign convention is that of Weinberg \cite{Weinberg1972}, where the 2-sphere has $R<0$. We could work equally well with the opposite sign convention, with the key point being that the sign of $R$ indicates different local geometry. Our thesis here and elsewhere is that attractive interactions correspond locally to the 2-sphere, and repulsive interactions correspond locally to the pseudo 2-sphere. $R$ has been physically interpreted with an extension of Einstein's thermodynamic fluctuation theory (see \cite{Ruppeiner2010} for a recent review). This extended theory includes covariance, conservation, and consistency, and clearly marks $|R|$ as a thermodynamic measure of the mesoscopic size scale where the character of fluctuations changes qualitatively due to fluctuating structures resulting from microscopic interactions. We point out that our discussion for two-parameter thermodynamic systems does not necessarily extend to three or more independent parameters. It has been argued \cite {Ruppeiner1998} that with three or more independent parameters, $R$ is still the dominant quantity. It is possible that our observations about the sign of $R$ might generalize beyond two-parameter systems, but there are as yet no good model calculations bearing on this point of view.}) \cite{Mrugala1990, Oshima1999}. These ideas were recently amplified in the condensed liquid and solid phases by the analysis of Lennard-Jones (LJ) computer simulation data, taken varying the interparticle interactions using the WCA perturbation ansatz \cite{May2013}. These authors also evaluated $R$ for the fcc-LJ solid, where uniformly positive $R$ was found. These recent studies make very natural an interpretation of positive $R$ in fluids (for which $R$ is mostly negative) as indicating solid-like properties.

\par
\begin{figure}
\centering
\includegraphics[width=7cm]{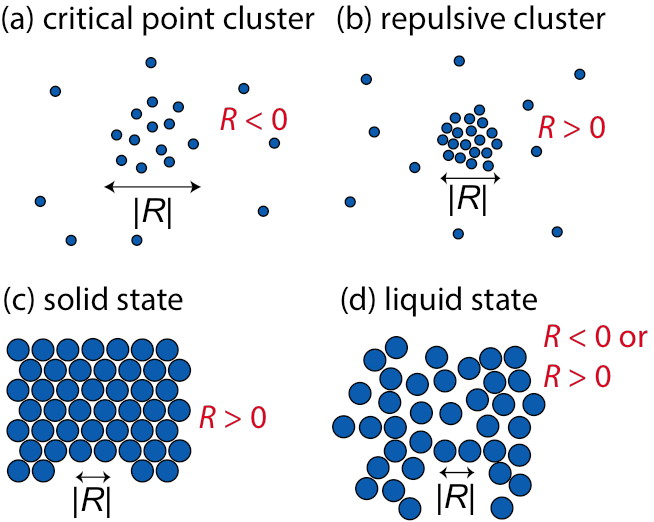}
\caption{Schematics of four organized particle arrangements and their $R$'s: (a) an attractive near critical point cluster, with $R<0$ and $|R|\sim\xi^3$; (b) a repulsive solid-like cluster held up by hard-core particle repulsion, with positive $R\sim$ cluster size; (c) the solid state, with positive $R\sim v$, and with $v$ the molecular volume; (d) the compact liquid state, with positive or negative $R$ and $|R|\sim v$.}
\label{fig:1}
\end{figure}

Figure \ref{fig:1} shows a schematic summary of various molecular configurations, and the associated sign of $R$ \cite{Ruppeiner2012a}. Key in this paper is the liquid state in Fig. \ref{fig:1}(d), where $R$ may have sign characteristic of the solid phase $R>0$, or the looser, more typical, liquid phase $R<0$. If the sign is positive we interpret this as solid-like liquid properties.

\par
The unique status of $R$ as the only geometric invariant in thermodynamics makes its evaluation important whenever possible. There is as yet no rigorous general proof connecting the sign of $R$ to the character of the microscopic interactions, so one must at present simply develop experience based on the evaluation of $R$ for many examples. The $R$-diagrams in this paper are an essential element of this effort.

\section{The calculation of $R$}

$R$ follows from the Helmholtz free energy per volume $f=f(T,\rho)$, with $T$ the temperature, and $\rho$ the number of particles per volume \cite{Ruppeiner2010}. Let $s=-f_{,T}$ be the entropy per volume, and $\mu=f_{,\rho}$ the chemical potential. The thermodynamic length is $\Delta\ell^2=g_{T}\Delta T^2 + g_{\rho}\Delta\rho^2$, with
 
\begin{equation} \left\{ g_{T}, g_{\rho}\right\} =\left\{\frac{1}{k_B T}\left(\frac{\partial s} {\partial T}\right)_{\rho},\frac{1}{k_B T}\left(\frac{\partial \mu} {\partial \rho}\right)_{T}\right\}, \label{10}\end{equation}

\noindent where $k_B$ is Boltzmann's constant. The thermodynamic curvature is

\begin{equation}
R=\frac{1}{\sqrt{g}}\left[\frac{\partial}{\partial T} \left(\frac{1}{\sqrt{g}}\frac{\partial g_{\rho}}{\partial T}\right)+\frac{\partial}{\partial\rho}\left(\frac{1}{\sqrt{g}}\frac{\partial g_{T}}{\partial \rho}\right)\right],\label{30}
\end{equation}

\noindent where $g=g_{T}\,g_{\rho}$. $R$ involves up to third derivatives of $f(T,\rho)$.

\par
\begin{figure}
%%%%%%%%%%%%%%%%%%%%%%%%%%%%%%
\begin{minipage}[b]{0.5\linewidth}
\includegraphics[width=2.7in]{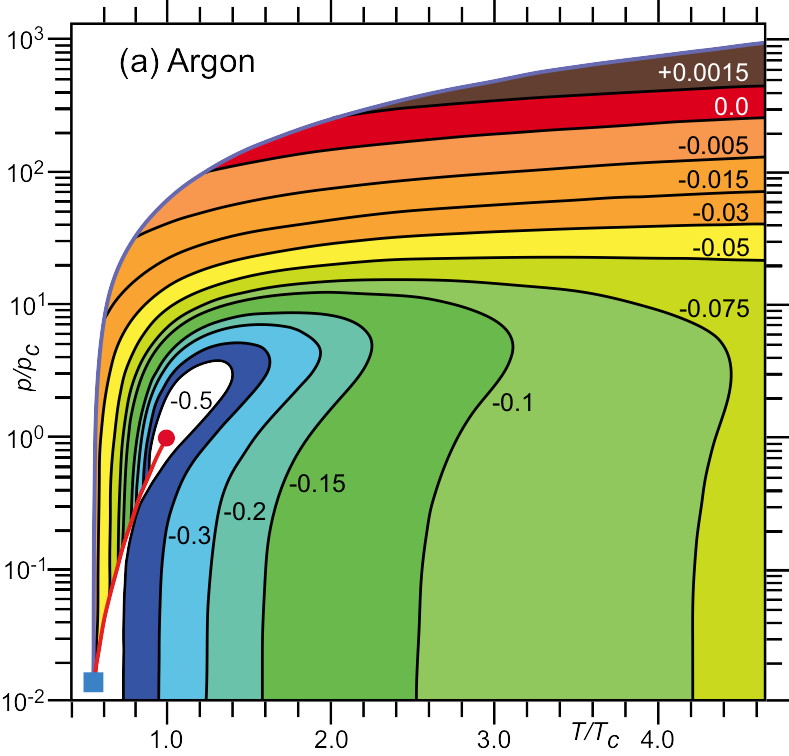}
\end{minipage}
\hspace{0.0 cm}
\begin{minipage}[b]{0.5\linewidth}
\includegraphics[width=2.7in]{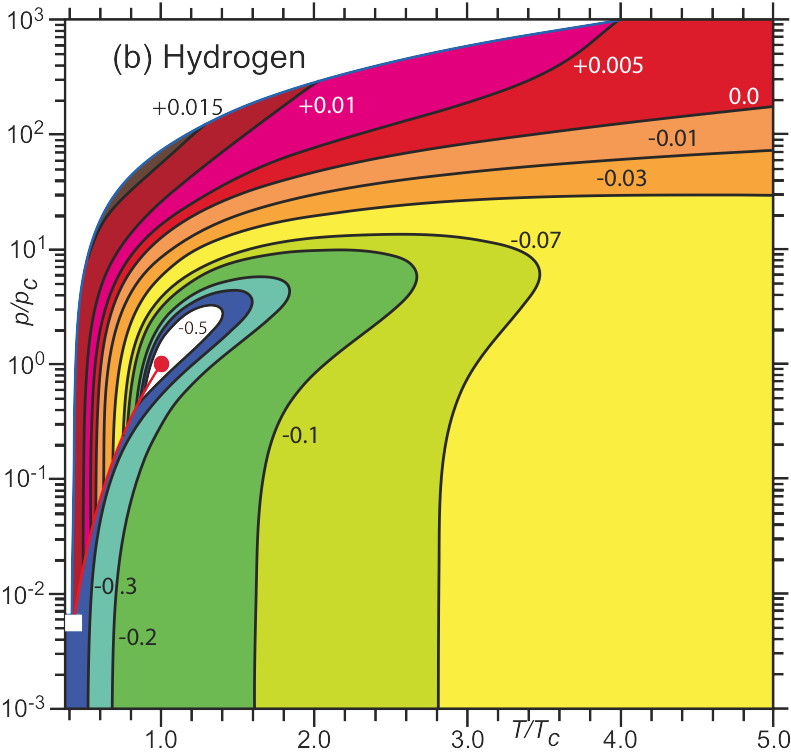}
\end{minipage}
\vspace{0.1cm}
%%%%%%%%%%%%%%%%%%%%%%%%%%%%%%%
\begin{minipage}[b]{0.5\linewidth}
\includegraphics[width=2.7in]{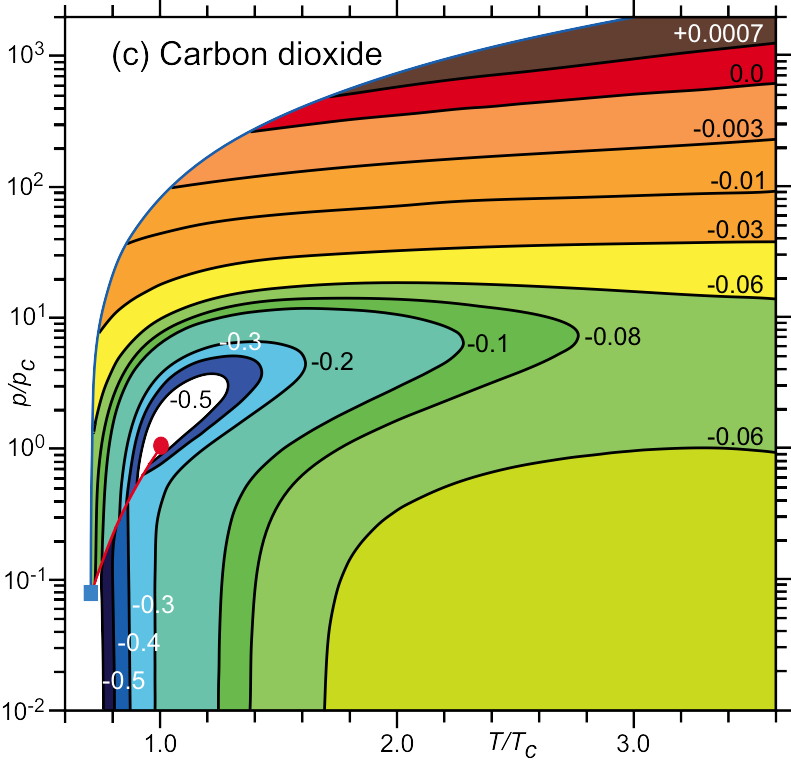}
\end{minipage}
\hspace{0.0 cm}
\begin{minipage}[b]{0.5\linewidth}
\includegraphics[width=2.7in]{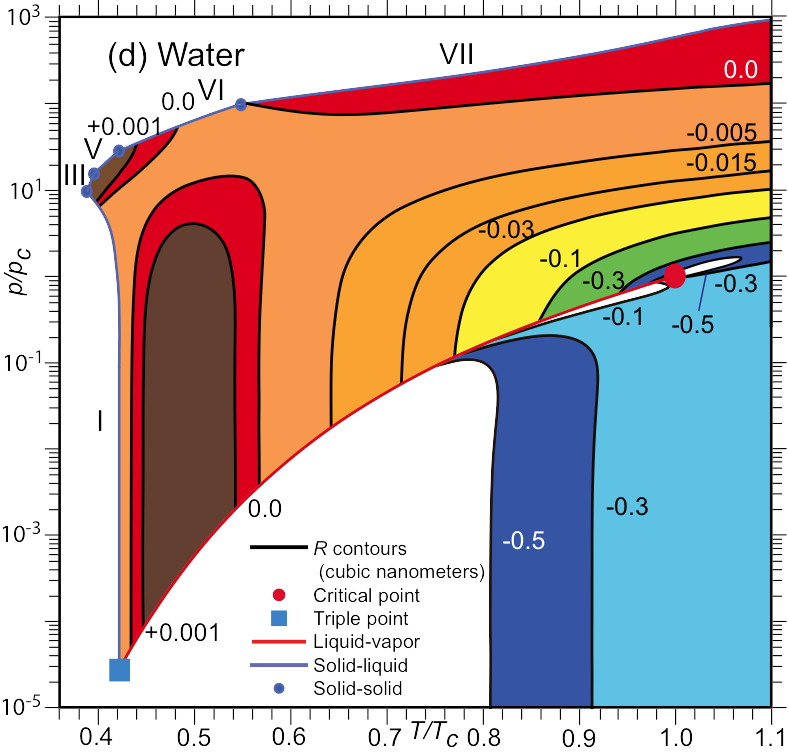}
\end{minipage}
%%%%%%%%%%%%%%%%%%%%%%%%%%%%%%%
\caption{$R$-diagrams for four fluids. The units of the $R$-contours are cubic nanometers, and we plot temperature $T$ and pressure $p$ in units of their critical values $T_c$ and $p_c$. Included in the plots are the triple point, the critical point, and the liquid-vapor and solid-liquid coexistence curves. Various solid phases in water are labelled.}
\label{fig:2}
\end{figure}

Phenomenological equations of state have been fit to experimental data for a number of fluids, with various critical temperatures $T_c$, and compiled by NIST \cite{NIST}. Such fits allow for the ready computation of $R$. Particularly good fits are those for argon ($T_c=150.69$ K) \cite{Tegeler1999}, hydrogen ($T_c=33.15$ K) \cite{Leachman}, carbon dioxide ($T_c=304.13$ K) \cite{Span1996}, and water ($T_c=647.10$ K) \cite{Wagner2002}, with corresponding contour plots of $R$ shown in Figure \ref{fig:2}. The fit accuracy for quantities involving second derivatives of $f(T,\rho)$ is typically about $1\%$, not too near the critical point. For the third derivatives of $f(T,\rho)$, fit accuracy is harder to gauge since direct experimental data are not available for comparison. However, the $R$'s constructed using these third derivatives are smooth functions, with values falling into a coherent picture. We expect the fit accuracy for calculating $R$ to be adequate.

\section{Results}

The $R$-diagrams in Fig. \ref{fig:2} display the following features: 1) In the supercritical liquid region near the melting line, there are significant areas of positive $R$, suggesting solid-like properties in the liquid. These areas are bounded below by $R=0$ contours. These positive $R$'s have values roughly the volume of an atom, and, for given molecular volume $v$, depend only weakly on $T$. Solid-like features in this regime have been identified in computer simulations, and are under active investigation \cite{Ryltsev2013}. 2) Towards lower temperatures, the $R =0$ contours usually, but not always, terminate on the solid-liquid coexistence curve. 3) Very near the critical point, the $R$-contours begin and end on the coexistence curve, and loop around the critical point, establishing a distinct, self-contained, critical point regime. 4) The $R$-contours take on increasingly negative values as we approach the critical point, at which $R$ diverges to negative infinity. 5) Further from the critical point, the $R$-contours start in the liquid phase, loop around the critical point, and exit the $R$-diagram at low pressure in the rarefied gas phase. Alternatively, the $R$-contours may start in the vapor phase, and exit the $R$-diagram at low pressure in the rarefied gas phase.

\par
\begin{figure}
%%%%%%%%%%%%%%%%%%%%%%%%%%%%%%
\begin{minipage}[b]{0.5\linewidth}
\includegraphics[width=2.7in]{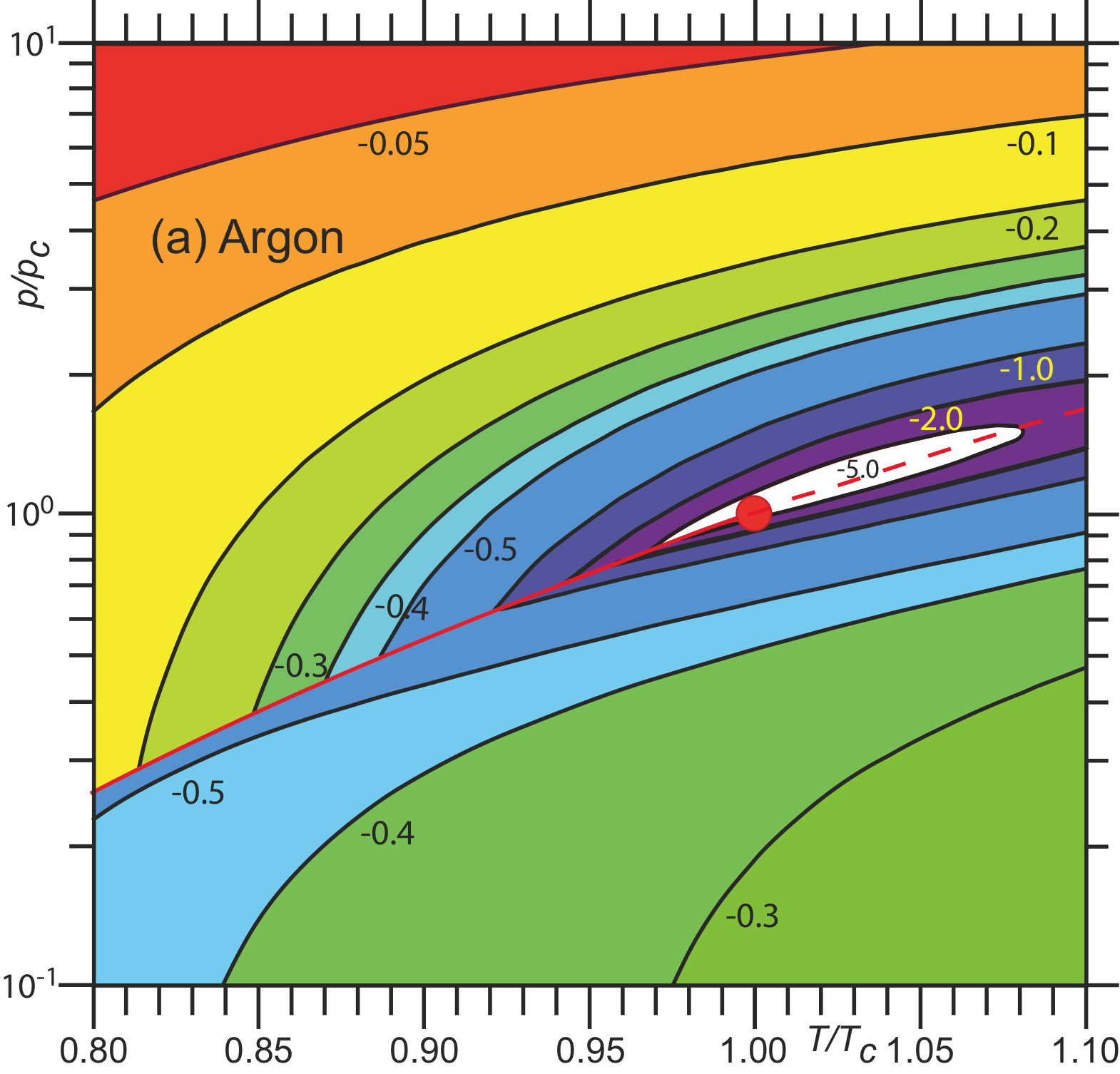}
\end{minipage}
\hspace{0.0 cm}
\begin{minipage}[b]{0.5\linewidth}
\includegraphics[width=2.7in]{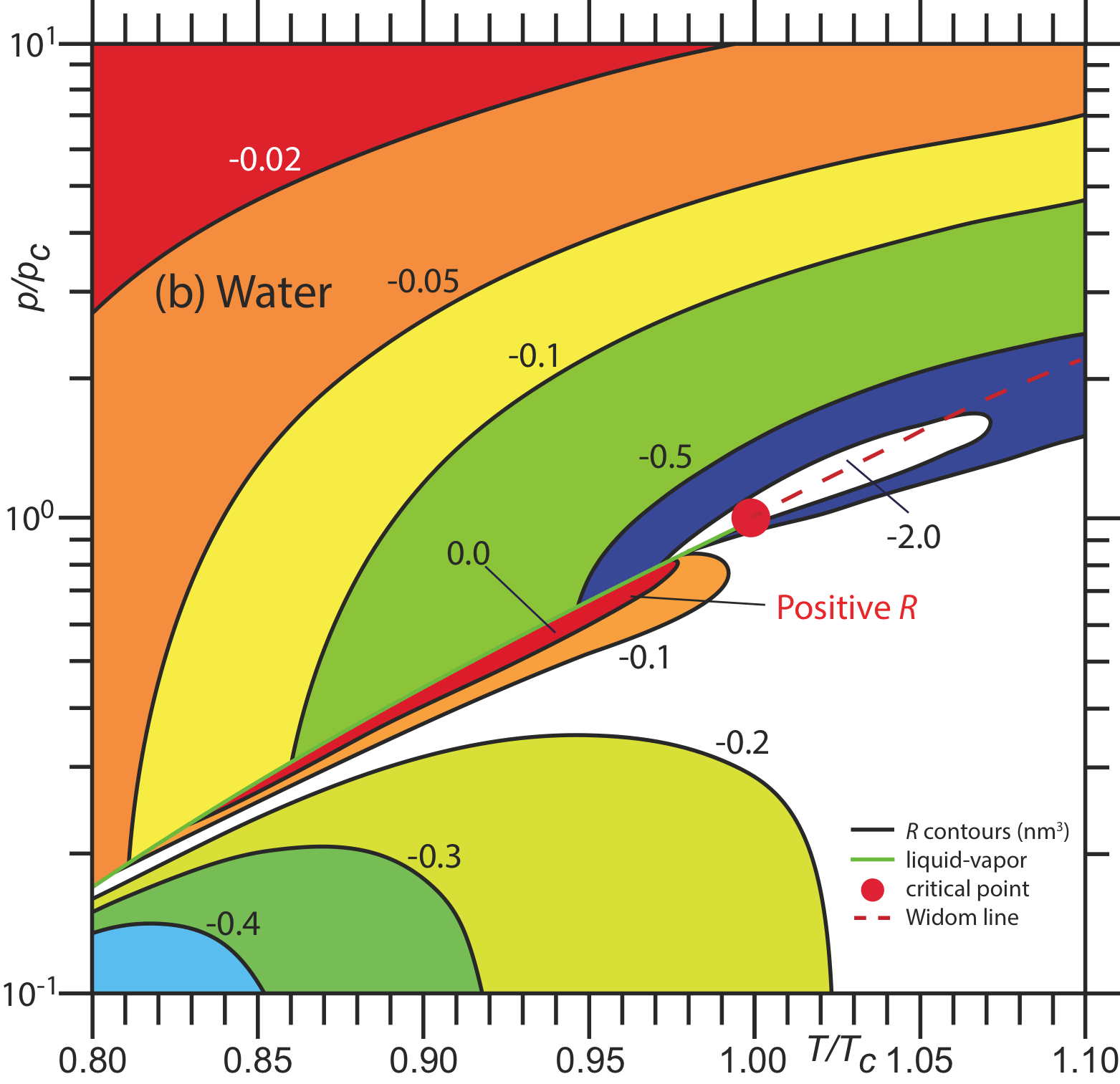}
\end{minipage}
%%%%%%%%%%%%%%%%%%%%%%%%%%%%%%%
\caption{Enlargement of the critical regimes of argon and water. Argon shows typical fluid properties, but water has a narrow region of positive $R$ in the gaseous state abutting the liquid-vapor coexistence curve. This region might reveal the presence of solid clusters in the gas.}
\label{fig:3}
\end{figure}

Water in Fig. \ref{fig:2}(d) shows some special features. In addition to the typical region of positive $R$ near the melting line of ice VII, our analysis reveals three more regions with positive $R$. First, there is a very narrow area on the vapor side of the liquid-vapor coexistence curve in the vicinity of the critical point; see Figure \ref{fig:3}(b) for an expanded view. (Argon is shown for comparison in Figure \ref{fig:3}(a), and with both fluids we display the Widom lines corresponding to the locations of maximum $|R|$ along lines of constant $p$ \cite{Ruppeiner2012b, May2012}.) This very narrow area of positive $R$ was first identified on the water coexistence curve, and was interpreted as indicating the presence of solid clusters in the vapor, of a type shown schematically in Fig. \ref{fig:1}(b) \cite{Ruppeiner2012a}. In addition, there is a narrow region with positive $R$ close to the melting line of ice III, V, and VI.

\par
Perhaps the most interesting region with positive $R$ in water is the large narrow slab-like area in the range $0.44 < T/T_c < 0.57$, starting at the liquid-vapor coexistence curve, and extending up in the liquid phase to pressure $p/p_c\sim 10$. In this region, $R$ peaks at about $0.002 \,\,\mbox{nm}^3$ along a vertical line with $T=315$ K, a value of $R$ consistent with the other solid-like values found here and elsewhere \cite{May2013}.

\par
Cold water is known for its anomalous behavior, including a density maximum at 4 C and ambient pressure. Anomalous behavior is present as well in the isothermal compressibility $K_T$, the isobaric heat capacity $C_p$, and the thermal expansion coefficient $\alpha_p$ \cite{Debenedetti2003b, Debenedetti2003a}. Both $K_T$ and $C_p$ increase as we cool isobarically into the metastable liquid state, while $\alpha_p$ decreases. The rates of change increase as the temperature is lowered. This indicates the approach to a possible singularity at sufficiently low temperature, in the form of a liquid-liquid critical point (LLCP) \cite{Poole1992}. Microscopically, the source of water's anomalies appear to be the presence of open tetrahedrally-coordinated networks of water molecules held together by hydrogen bonds (HB)\cite{Debenedetti2003b}. Computer simulations contribute in a key way to this picture \cite{Tanaka2012, Limmer2013, Palmer2014}.

\par
In the stable phase, this picture of water connects directly to our $R$-diagram in Fig. \ref{fig:2}(d). In recent studies of experimental data, Mallamace {\it et al.} \cite{Mallamace2013, Mallamace2012} analyzed the $(p,T)$ dependence of several thermodynamic response functions to find limits of regions with anomalous behavior. These authors reported that the isobaric density maximum is strongly $p$-dependent, and that it disappears above a certain crossover pressure $p_{cross} \sim 8.1 p_c$, coinciding closely with the top of our positive $R$ narrow slab region in Fig. \ref{fig:2}(d). Mallamace {\it et al.} also noted that the minimum in $K_T$ is nearly $p$-independent, and occurs at a temperature $T^*=0.49\,T_c$, coinciding almost exactly with the central vertical line of our narrow slab. Mallamace {\it et al.} identified as well some other special features of $T^*$ and concluded that the ``singular and universal expansivity point'' \cite{Mallamace2012} at $T^*$, and the crossover pressure $p_{cross}$, characterize the thermodynamic and structural properties of water. In addition, they suggested that $T^*$ may mark the onset of HB clustering with ice-like low-density liquid (LDL) structures more compressible than the high-density liquid (HDL) structures at higher $T$. This certainly conforms to our interpretation of positive $R$ in liquids. We add that, in addition to revealing the top and center line of this region of interest at a glance, our water $R$-diagram also reveals the upper and lower bounding temperatures.

\par
Our results also relate to a possible second critical point in water. Starting in the narrow slab region, and lowering $T$ at constant $p$ into the supercooled liquid state, leads to negative and decreasing $R$. This would be expected if there were a second critical point, since all of the known fluid critical point models tabulated in \cite{Ruppeiner2010} have $R$ decreasing to negative infinity at the critical point. The LLCP is expected to be in the same universality class as the liquid-vapor critical point \cite{Holten2012}. Clearly then, the narrow slab of positive $R$ must terminate at some lower value for $T$, and this is indicated in the $R$-diagram Fig. \ref{fig:2}(d). Water is the only one of the four fluids in Fig. \ref{fig:2} with $R$ decreasing on cooling into the metastable liquid regime. Recently, Holten {\it et al.} \cite{Holten2012} analyzed the behavior of metastable supercooled water assuming a LLCP. We are in the process of making a detailed analysis of $R$ in this special region, and have confirmed that $R$ indeed diverges to negative infinity at the theoretical LLCP.

\par
It is of some interest to examine our $R=0$ line in the supercritical state for compliance with other theoretical lines in this region, e.g., the Frenkel line \cite{Bolmatov2013}, or the Fisher-Widom line \cite{Fisher1969}. We could not find any correspondence with the Frenkel line. A comparison with the Fisher-Widom line is difficult because the correct determination of this line is still an unsolved research issue.

\section{Conclusion}
In conclusion, we have presented a new tool for viewing fluid properties, the $R$-diagram, which reveals a number of features which are otherwise difficult to see. We presented $R$-diagrams for four fluids, and highlighted regions of positive $R$, which we propose correspond in the liquid phase to solid-like features. Water was found to be particularly interesting in this respect. Since $R$ is the only thermodynamic geometric scalar invariant, it seems important to examine it for any system with known thermodynamics.

\section{Acknowledgements}

GR thanks George Skestos for research support, Horst Meyer for help with the melting curve for hydrogen, and INFN in Frascati, Italy, where some of this work was written in final form, for their hospitality.

\end{document}